# Ultrafast helicity control of surface currents in topological insulators with near-unity fidelity

Christoph Kastl[1], Christoph Karnetzky[1], Helmut Karl[2], Alexander W. Holleitner[1*]

[1] *Walter Schottky Institut and Physik-Department, Technische Universität München, Am Coulombwall 4a, 85748 Garching, Germany.*

[2] *Institute of Physics, University of Augsburg, 86135 Augsburg, Germany.*

**In recent years, a class of solid state materials, called three-dimensional topological insulators, has emerged. In the bulk, a topological insulator behaves like an ordinary insulator with a band gap. At the surface, conducting gapless states exist showing remarkable properties such as helical Dirac dispersion and suppression of backscattering of spin-polarized charge carriers. The characterization and control of the surface states via transport experiments is often hindered by residual bulk contributions yet at cryogenic temperatures. Here, we show that surface currents in $Bi_2Se_3$ can be controlled by circularly polarized light on a picosecond time scale with a fidelity near unity even at room temperature. We reveal the temporal separation of such ultrafast helicity-dependent surface currents from photo-induced thermoelectric and drift currents in the bulk. Our results uncover the functionality of ultrafast optoelectronic devices based on surface currents in topological insulators.**

Layered materials, such as $Bi_2Te_3$[1],[2], Sb[3], $Bi_2Se_3$[2],[4],[5],[6], $Bi_2Te_2Se$[7], and $(Bi_{1-x}Sb_x)_2Te_3$[8], are important narrow-bandgap semiconductors for tunable, high-performance infrared detectors and thermoelectric applications[9]. They have been demonstrated to be reference three-dimensional topological insulators[10][11][12] exhibiting exceptional transport mobilities[1][2][3][4][5][6]. The latter suggest a reduced energy consumption which is very attractive for semiconductor devices in high-speed communication applications. In this respect, it is very advantageous that the helical surface states in topological insulators can be addressed by polarized light[2][9][13][14][15][16]. Particularly, the circular photogalvanic effect results from a helicity-dependent asymmetric optical excitation of spin-split surface states in momentum space, allowing the generation and control of spin-polarized surface currents in topological insulators by circularly polarized light[17][18][19]. Here, we demonstrate that such surface currents can be accessed and read-out independently of the bulk photocurrents on a picosecond time scale with near-unity fidelity. Therefore, our results open the avenue for a high-speed transmission of information based on topological insulators.

The investigated n-type $Bi_2Se_3$-films are embedded in a metal-topological insulator-metal photodetector geometry (Fig. 1a). We characterize the ultrafast photocurrents by an on-chip, time-domain THz photocurrent spectroscopy with a picosecond time resolution (Fig. 1b and Methods)[20][21][22]. We excite the $Bi_2Se_3$-films with a circularly polarized pump pulse under an oblique angle $\theta$ (Fig. 1b). Such helical photons excite spins asymmetrically in $k$-space owing to angular momentum selection rules in $Bi_2Se_3$[9][23]. In turn, a net out-of-equilibrium spin polarization is acquired in the Dirac cone. After photoexcitation, spin and charge degrees of freedom relax on different time scales in the bulk and surface states of $Bi_2Se_3$[24]. For the surface states, spin depolarization, intraband cooling via surface electron-phonon scattering, and interband electron-hole recombination occur on a sub-picosecond to picosecond time scale [24][25][26].

Figure 1c shows an image of a $Bi_2Se_3$-film with two gold striplines serving as electronic contacts. We scan the pump laser across such a $Bi_2Se_3$-film and record the time-integrated photocurrent $I_{photo}$ as a function of the laser position (Fig. 2a). At the metal interfaces (triangles), a photothermoelectric current is generated because of a laser-induced heat gradient and the large thermoelectric power of $Bi_2Se_3$. In between the two striplines, the time-integrated photocurrent averages to be close to zero (circles in Figs. 2a and 2b). Changing the polarization of the exciting laser from circularly left to horizontally linear and then to circularly right (Fig. 2c), we observe the sinusoidal fingerprint[15] of a circular photogalvanic current with an amplitude '$C$' (Methods and Supplementary Fig. 1). The sinusoidal fits consider the circular and linear photogalvanic currents depending on the oblique angle $\theta = +17°$ (Supplementary Note 1 and Supplementary Fig. 1). The different background amplitudes of $I_{photo}$ are highlighted as dashed lines in Fig. 2c. They represent the spatially varying, dominant photothermoelectric current $I_{thermo}$. We introduce a fidelity $f = C / (C + I_{thermo})$, which describes the ability to resolve the circular photogalvanic current with respect to $I_{thermo}$. For all samples, we find $f$ to be smaller than 40% in time-integrated measurements. The maximum value is achieved for the laser being positioned at the center of the $Bi_2Se_3$-film (e.g. circles in Fig. 2), where the overall $I_{thermo}$ generated within the laser spot averages out towards the noise amplitude ($I_{thermo} \rightarrow A_{noise} \sim 40$ pA). As we show below, high-speed, time-resolved measurements provide a fidelity close to unity. This high fidelity can be achieved because photocurrent contributions on different time scales with different directions do not average out.

We measure the time-resolved photocurrent $I_{sampling}$ for different excitation positions on a $Bi_2Se_3$-film from the right to the left contact (right- and leftward triangles in Fig. 3a). For all positions, we observe that $I_{sampling}$ changes sign and therefore direction at a certain $\Delta t$. This means that for short (long) times, a current with a direction towards the closer (farther) contact dominates. The origin of the opposing currents can be understood as follows. Within the first picosecond after excitation, the electrons thermalize to form a hot carrier ensemble[27],[28]. They propagate away from the laser spot due to sample-internal potentials such as the thermopower and the density gradient. Close to the metal-interfaces, the thermopower generated between $Bi_2Se_3$ and Au gives rise to the photothermoelectric current.

In a time-of-flight analysis, we determine the ultrafast transport currents with opposing directions for each laser position (Supplementary Figures 2 and 3). The analysis allows us to estimate the fastest time-of-flight velocity of photogenerated hot electrons to be $v_e = (5.7 \pm 1.5) \times 10^5$ ms$^{-1}$ at room temperature (Supplementary Note 2) which agrees remarkably well with the group velocity of the $Bi_2Se_3$ at the given Fermi-energy[29][30], as discussed below. In addition, we determine the decay time of the photocurrent signals to be $(3.0 \pm 0.5)$ ps at room temperature which is in agreement with the picosecond relaxation dynamics of hot electrons[26][27].

In a next step, we focus the laser onto the center of the $Bi_2Se_3$-film (Fig. 3b) where the time-integrated photocurrent is close to zero (e.g. circle in Fig. 2a). At high bias, the data mimic the curves measured at zero bias at the right and left contacts (right- and leftward triangles in Fig. 3a). This means that at the center in between the striplines (Fig. 3b), an externally applied $V_{sd}$ gives rise to an additional drift of the photogenerated hot electrons. The polarity of the applied bias allows us to confirm that the dominating contribution stems from electrons and not holes (Supplementary Note 2 and Supplementary Fig. 4). The similarity of the top and bottom traces in Figs. 3a and 3b suggests that for certain values of $V_{sd}$ at the center, the corresponding electrostatic potential has an equivalent impact on the hot electron dynamics as the thermopower at the metal-interfaces. In turn, at the center of the sample and for $V_{sd} = 0$ V, the net field due to sample-internal potentials is zero. Particularly, the temporal tails of $I_{sampling}$ are close to zero for $\Delta t \geq 10$ ps at zero bias. We note that there is always an ultrafast photocurrent within the first picoseconds at this position (Fig. 3b).

Peculiarly, the sign and amplitude of the prevailing ultrafast contribution are completely controlled by the polarization of the exciting photons and their oblique angle of incidence. Figure 4a shows $I_{sampling}(\Delta t)$ measured close to the center of another $Bi_2Se_3$-film for a varying photon-polarization. The fitting curves consider ultrafast transport currents with changing directions depending on the photon-polarization (Supplementary Fig. 2). Particularly, for $\Delta t \approx 4$ ps, the current amplitude depends only on the photon-polarization. Microscopically, the helicity dependent contribution results from an asymmetric excitation in $k$-space of the spin-polarized surface states caused by the peculiar helical symmetry of the surface states (Supplementary Fig. 1). The panels in Fig. 4a sketch the direction of the helicity-dependent current for circularly left and right polarized light (red and blue). The corresponding, measured peak amplitude follows the sinusoidal fingerprint of the circular and linear photogalvanic effects (Fig. 4b) with a fidelity $f$ exceeding 95% for all temperatures. At the same ex-citation position, when the polarization dependence is measured in a time-integrated manner (Fig. 4c), the time-averaged background reduces $f$ to be 32.6%, because the photocurrent contributions with different directions compete on long time scales.

It is insightful to highlight the dynamics of the photocurrents in Fig. 4a again. After the spin depolarization and intraband cooling ($\Delta t \geq 5$ ps), the spin-information and helicity protection is predominantly lost. Then, the expansion dynamics of the same (hot) electrons dominate the transport dynamics. These currents follow the combined influence of the thermopower and the electrostatic potentials. In the following, we will derive further insights into the interplay of the two potentials. Starting point is the fact that the optoelectronic expansion dynamics are dominated by photogenerated hot electrons which propagate at $v_e = (5.7 \pm 1.5) \times 10^5$ ms$^{-1}$ (Supplementary Note 2). Generally, the dispersion of the surface states can be written as[31]

$$E_{\pm}(k) = E_0 - Dk^2 \pm \sqrt{\left(\frac{\Delta}{2} - Bk^2\right)^2 + (v_{\text{Dirac}}\hbar k)^2} \qquad (1),$$

with $\hbar k$ the in-plane crystal momentum, $v_{\text{Dirac}}$ is group velocity close to the Dirac point, $D = -12.4$ eVÅ$^2$ a quadratic term resulting from the broken particle-hole symmetry[32], $B = 0$ eVÅ$^2$ describing massive states[32], and $\Delta$ the energy gap for the inter-surface coupling which is zero for the investigated samples. The value for the binding energy of the Dirac point $E_0 = (0.55 \pm 0.05)$ eV is determined from Hall-measurements (assuming an effective mass of $0.13 \cdot m_e$)[33]. Recent photoemission experiments on $n$-type Bi$_2$Se$_3$ and related topological materials demonstrate that only a few 100 fs after photoexcitation, the electron distribution is already centered at the Fermi-energy[25], [27], [28]. From equation (1), we derive the group velocity $v_{\text{group}} = \frac{1}{\hbar}\frac{\partial E}{\partial k}$ to be $(5.5 \pm 0.2) \times 10^5$ ms$^{-1}$ at the Fermi-energy $E_{\text{Fermi}}$. We note that for bulk states, the principal order of the group velocity is in agreement with this value at the given Fermi-energy[32]. Importantly, the experimental value of $v_e$ agrees remarkably well with the derived value and with the reported one determined from photoemission experiments[29][30]. All values exceed the saturation drift velocity as measured in $Bi_2Se_3$-based transistor devices[34]. In other words, the group velocity of photogenerated charge carriers determines the ultimate speed of the optoelectronic response in topological insulators.

Experimentally, this manifests itself in a linear time-of-flight diagram with $v_e = (5.7 \pm 1.5) \times 10^5$ ms$^{-1}$ (Supplementary Figures 2 and 3). To further discuss this surprising result, we use a two-temperature model $T_e$ and $T_{phonon}$ for the electron and phonon baths[35][36]. The electron heat capacity is well approximated by a linear temperature dependence $C_e \sim C'_e \cdot T_e$, if $T_e \ll T_{Fermi}$ with $T_{Fermi} = 3320$ K the Fermi-temperature of the measured n-type $Bi_2Se_3$-films[37]. Within this model, we calculate the maximum electron temperature $T_e^{max} = 1500$ K for the experimental parameters as in Fig. 3[35]. The electron-electron collision time can be estimated to be $\tau_{ee} \sim \hbar \cdot E_{Fermi} / (k_B \cdot T_e)^2 = 9$ fs, with $k_B$ the Boltzmann constant[37], and the electron-phonon scattering time is on the order of $\tau_{e\text{-}phonon} \sim \hbar / (k_B \cdot T_{bath}) = 94$ fs. The latter was recently determined to be ~0.7 ps for optical phonons and ~2.3 ps for acoustic phonons[24],[27]. In this regime ($\tau_{ee} \ll \tau_{e\text{-}phonon}$), there exists a strong electron-lattice non-equilibrium, and the electron relaxation is governed by the electron-electron collisions[35]. It was reported that hereby, the electronic heat transport occurs at the Fermi velocity after an ultrafast optical excitation of the electron bath[36]. The dynamics can be analytically described by a two-temperature model in the limit of a strong electron-lattice non-equilibrium in combination with an electron heat conductance $\kappa_e \propto (T_e)^{-1}$ [35]. In the present experiment on $Bi_2Se_3$-films, the ultrafast current of hot electrons is measured. In our inter-pretation, this current carries the electronic heat to the contacts, and the corresponding heat transport at the Fermi velocity explains our data with linear time-of-flight diagrams (Supplementary Figures 2 and 3).

Our on-chip, time-domain THz spectroscopy allows us to reveal the impact of an electrostatic potential on such non-equilibrium expansion dynamics (Figure 3b). Along this line, we estimate the maximum gradient of the local electron temperature profile to be $\nabla T_e^{max} \sim T_e^{max} / (0.5 \cdot d_{stripline}) = 1500$ K $/ (0.5 \cdot 15$ μm$) = 200$ Kμm$^{-1}$, with $d_{stripline}$ the distance between the two striplines. In first approximation, the measured expansion dynamics of the hot electrons are identical for exciting the Bi$_2$Se$_3$-films at the contacts (top and bottom traces in Figs. 3a) and for an excitation spot in the center of the sample at finite bias (top and bottom traces in Fig. 3b). For the results in Fig. 3b, the laser position is carefully chosen such that influence of the thermopower induced at the contacts is close to zero. Then for a finite bias $V_{sd}$, the electrostatic potential gradient $\nabla V_{electrostatic}$ at the laser spot has an equivalent impact on the hot electron dynamics as the thermopower at the metal-interfaces. We describe this thermopower at very short time scales as the product of $S_{Bi2Se3} \cdot \nabla T_e$, with $S_{Bi2Se3}$ a non-equilibrium, effective Seebeck coefficient. Comparing the amplitudes of the top and bottom traces of Figs. 3a and 3b, we extract $|\nabla V_{electrostatic}| = (820 \pm 440)$ Vm$^{-1}$ (Methods). In turn, we can estimate the effective Seebeck-coefficient to be in the order of $S_{Bi2Se3}^{minimum} \sim \nabla V_{electrostatic} / \nabla T_e^{max} = -(4.1 \pm 2.2)$ μVK$^{-1}$. The extracted value phenomenologically describes the non-equilibrium thermopower at the Bi$_2$Se$_3$-metal contacts. To the best of our knowledge, the above derivation is the first estimate of an effective Seebeck-coefficient of non-equilibrium hot electron ensembles after a pulsed laser excitation. The derived non-equilibrium Seebeck-coefficient is smaller than the typical quasi-equilibrium value $S \sim -50$ μVK$^{-1}$ [38]. This can be understood as follows. For the exper-imental parameters in Fig. 3, we concurrently find a time-averaged amplitude $|I_{photo}|$ = 370 pA at the $Bi_2Se_3$-metal contact and at an acquisition time of ~ms. At this long time-scale, the heat transport is governed by phonons. Accordingly, we numerically calculate a temperature increase at a $Bi_2Se_3$-metal contact of $\Delta T$ = 47 mK (for both the phonon and electron baths). In turn, we derive a quasi-static Seebeck-coefficient by the following expression $\Delta V \sim S_{Bi2Se3}^{quasistatic} \cdot \Delta T$, with $S_{Bi2Se3}^{quasistatic}$ = -23 µVK$^{-1}$. For the sample and experimental conditions as in Figure 2, we calculate $S_{Bi2Se3}^{quasistatic}$ = -62µVK$^{-1}$ at room temperature. Both values are consistent with the above quasi-equilibrium value of the Seebeck coefficient. In other words, at long time-scales, the heat transport is dominated by phonons. However, at ultrashort time-scales, it is governed by a highly non-equilibrium expansion of hot electrons. In our interpretation, this non-equilibrium current of hot electrons carries the electronic heat to the contacts as long as there exists a strong electron-lattice non-equilibrium as discussed above. Intriguingly, despite of such high-speed dynamics of the hot electrons, the data at $\Delta t$ ~ 4 ps in Fig. 4a demonstrate that the helicity-dependent currents can be still addressed and read-out. The underlying physical reason is the protection of the helical states within the spin depolarization time. Only for $\Delta t \geq 5$ ps in Fig. 4a, all currents have the same negative sign. That means that for this particular experiment, still a tiny thermopower potential drives these currents to the left contact, which is consistent with the negative offset observed in the time-integrated measurement (Fig. 4c). Again, for $\Delta t \geq 5$ ps, there is no polarization control for these currents anymore, since the pump-laser pulse is off and the spin depolarization has occurred (Fig. 4a).

The circular and linear photogalvanic effects are induced via surface states and not bulk states because of the broken inversion symmetry at the surface of $Bi_2Se_3$[15],[17]. A normal incidence geometry ($\theta = 0°$) suppresses the photogalvanic effects (Supplemen-tary Fig. 1), indicating an in-plane spin distribution and more fundamentally, an in-plane rotational symmetry of the involved electron states in $Bi_2Se_3$. Generally, there may be a small contribution from a helicity-independent transverse photon drag effect[17]. Contributions from Rashba-split bulk states at a surface inversion layer can be assumed to be negligible for the examined range of $E_{photon}$ (Supplementary Fig. 5). Overall, the ultrafast polarization-controlled photocurrent is limited by the spin lifetime in $Bi_2Se_3$.

To conclude, our experiments demonstrate the temporal separation of helicity-dependent surface photocurrents from polarization-independent bulk currents even at room temperature, and they reveal the onset of photo-thermoelectric currents at ultrafast time-scales. We elaborate the connection between the non-equilibrium currents of hot electrons directly after the pulsed laser excitation and the time-averaged thermoelectric currents which are typically described in the frame-work of a Seebeck coefficient. A time-of-flight analysis yields a speed of the photogenerated electrons which is consistent with the group velocity at the Fermi-energy of the $Bi_2Se_3$. Our results significantly advance the understanding of the optical excitation scheme and the ultrafast photocurrent dynamics in topological insulators. The time-delayed photothermoelectric currents in $Bi_2Se_3$ are partly caused by the n-doping of the material. They can be reduced by utilizing topological insulators with a Fermi-energy in the Dirac-cone, such as $Bi_2Te_2Se$ or $(Bi_{1-x}Sb_x)_2Te_3$[7],[8]. The picosecond response time of the surface currents proves the anticipated potential of topological insulators as promising materials for high-speed optoelectronic applications from the THz to the infrared range.

**Fabrication of the $Bi_2Se_3$-films.** The investigated $Bi_2Se_3$-films have a thickness of 50 nm ≤ 250 nm and a lateral dimension of more than 15 μm. We produce the $Bi_2Se_3$-films from 99.999 % pure crystalline $Bi_2Se_3$ granulate via exfoliation. We analyze the homogeneity and the dimensions of the films using optical microscopy, atomic force microscopy, and white light interferometry. The presented results have been reproduced on three independent samples, which have a thickness of 75 nm, 90 nm, and 150 nm. Hall-measurements on a 65 nm thin $Bi_2Se_3$-film from the same batch yield an electron density of ~4·$10^{19}$ cm$^{-3}$ at room temperature.

**Design of the stripline circuit.** Starting point is a 430 μm thick sapphire substrate covered with a 300 nm thin layer of ion-implanted silicon (Si). In a first optical lithography step, the Auston switch geometry is formed via HF-etching. The remaining silicon strip serves as a field probe (Auston switch). $Bi_2Se_3$-films are mechanically transferred onto the substrate by exfoliation. In a second optical lithography step, we evaporate 10 nm titanium (Ti) and 110 nm - 200 nm gold (Au) to form the waveguide circuits contacting the $Bi_2Se_3$-films and the read-out of the Auston switch. The distance between two parallel striplines is 15 μm. Each stripline itself has an individual width of 5 μm. The $Bi_2Se_3$-films are placed at a typical distance of 100 μm ≤ 800 μm from the Auston switch. The bias voltage $V_{sd}$ is applied between striplines (see Fig. 1). The striplines including the $Bi_2Se_3$-films have a two-terminal resistance of 2.9 kΩ, while the $Bi_2Se_3$-films have a resistance of ~55 Ω. Hereby, the gradient $\nabla V_{electrostatic}$ can be estimated from the bias voltage via $\nabla V_{electrostatic} = V_{sd}$ / 15 μm · 55 Ω / 2.9 kΩ.

**Time-integrated photocurrent spectroscopy** The time-averaged $I_{photo}$ is measured in between the two striplines (Fig. 1b) with a current-voltage amplifier.

**On chip, time-domain THz photocurrent spectroscopy.** The $Bi_2Se_3$-films in the stripline circuit are optically excited by a pump-pulse with ~200 fs pulse length generated by a titanium:sapphire laser at a repetition frequency of ~76 MHz with a photon energy of $E_{photon}$ = 1.53 eV and $E_{photon}$ = 1.59 eV. After excitation, an electro-magnetic pulse starts to travel along the striplines. A field probe senses the transient electric field of the travelling pulse (Fig. 1b). Here, we utilize an Auston-switch based on ion-implanted silicon. At a time delay $\Delta t$ with respect to the pump-pulse, the Auston-switch is short-circuit by a probe-pulse for the duration of the lifetime of the photogenerated charge-carriers in the silicon ($\tau \leq$ 1 ps). During this time-period, the transient electric field present at the field probe drives the current $I_{sampling}$. In turn, measuring $I_{sampling}(\Delta t)$ yields information on the optoelectronic response of the $Bi_2Se_3$ with a picosecond time-resolution. The time a THz-pulse travels from the $Bi_2Se_3$-film to the field probe can be estimated via $t_{travel} = d \cdot n_{sapphire}$ /c = 0.5 mm · 3.07 /c ~ 5.1 ps for a distance of 0.5 mm between the $Bi_2Se_3$-film and the Auston switch. The striplines have a total length exceeding 48 mm. Thus, reflections at the end of the striplines are expected for $\Delta t \geq$ 490 ps. These reflections are strongly reduced due to damping and radiation losses. Therefore, no reflections overlap with the electro-magnetic signal coming directly from the $Bi_2Se_3$-films in our data. The data of Figures 2 and 4 are presented for room temperature. The time-of-flight data in Figure 3 are depicted for 77 K, because the transfer characteristics of the striplines and therefore the signal-to-noise ratio are enhanced at lower temperatures. A time-of-flight analysis of data at room temperature is presented in the Supplementary Fig. 3. The position of the pump-spot is set with a spatial resolution of ~100 nm, while the position of the probe-spot is kept constant throughout the experiments. All measurements of $I_{sampling}$ were carried out utilizing an optical chopper system, a current-voltage-converter connected to the field-probe, and a lock-in amplifier. The spot size (FWHM) of the pump-laser is 3 - 4 μm for all time-resolved experiments and up to 9 μm for the time-integrated experiments. The laser power of the pump (probe) pulse is chosen to be in the range of 0.1 mW – 20 mW (80 mW – 150 mW). All data are taken in vacuum (~$10^{-5}$ mbar) to prevent the effect of photo-desorption of oxygen on the surface of the $Bi_2Se_3$. The measurements have been reproduced for different temperatures between 4 K and 295 K.

**Polarization control and sinusoidal fitting function.** The polarization-dependent photocurrents are characterized by measuring $I_{photo}$ or $I_{sampling}$, while rotating a $\lambda$/4-waveplate by an angle $\phi$. The rotation changes the photon-polarization with a period of 180° from linearly polarized ($\phi$ = 0°) to left circular ($\phi$ = 45°), to linearly ($\phi$ = 90°), to right circular ($\phi$ = 135°), and to linearly ($\phi$ = 180°). The fitting curves in Fig. 2c, 4b, and 4c, describe the photocurrent in the y-direction of the $Bi_2Se_3$-films (Fig. 1b) as $j(\phi)$ = C sin 2$\phi$ + $L_1$ sin 4$\phi$ + $L_2$ cos 4$\phi$ + D. As recently demonstrated[15], C describes the helicity-dependent circular photogalvanic effect with a rotational in-plane symmetry in $Bi_2Se_3$. $L_1$ comprises the helicity-independent linear photogalvanic effect. $L_2$ and D are bulk contributions (Supplementary Note 1).

We thank Leonid Levitov and Michael Knap for insightful discussions, as well as Andreas Brenneis and Ursula Wurstbauer for technical assistance. This work was supported by the DFG via SPP 1666 (grant HO 3324/8), ERC Grant NanoREAL (n°306754), the DFG excellence cluster "Nanosystems initiative Munich (NIM)" and the "Center of NanoScience (CeNS)" in Munich.

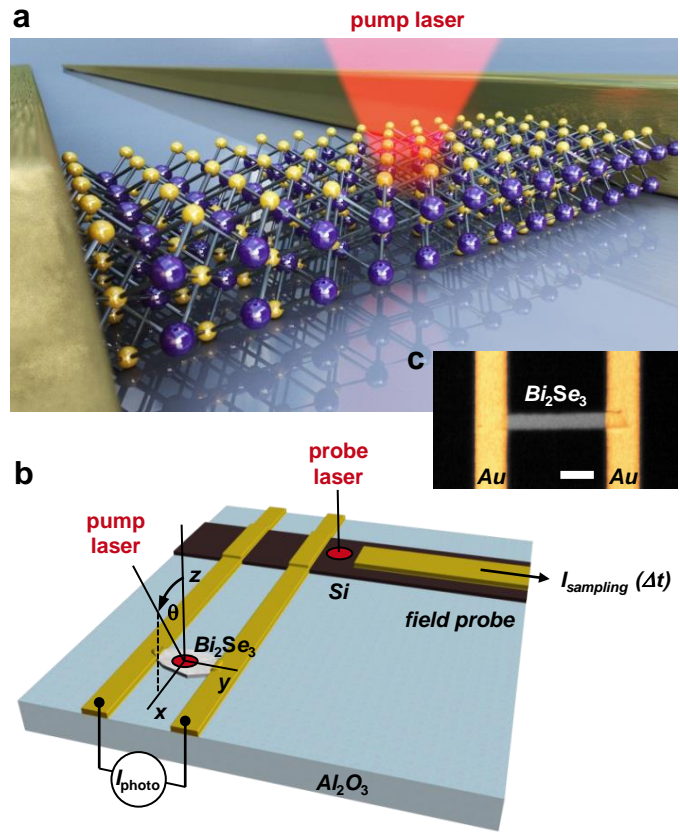

Fig. 1. Ultrafast photodetector based on topological insulators. (a) Circularly polarized photons of the pump pulse control surface currents in $Bi_2Se_3$. (b) The time-resolved photocurrent $I_{sampling}$ is read-out by a silicon-based Auston switch triggered by a probe pulse at a time-delay $\Delta t$. The time-integrated photocurrent $I_{photo}$ is measured between the electronic contacts which form two co-planar striplines. (c) Optical microscope image of a $Bi_2Se_3$-film with 75 nm thickness, contacted by two Au striplines. Scale bar, 5 µm.

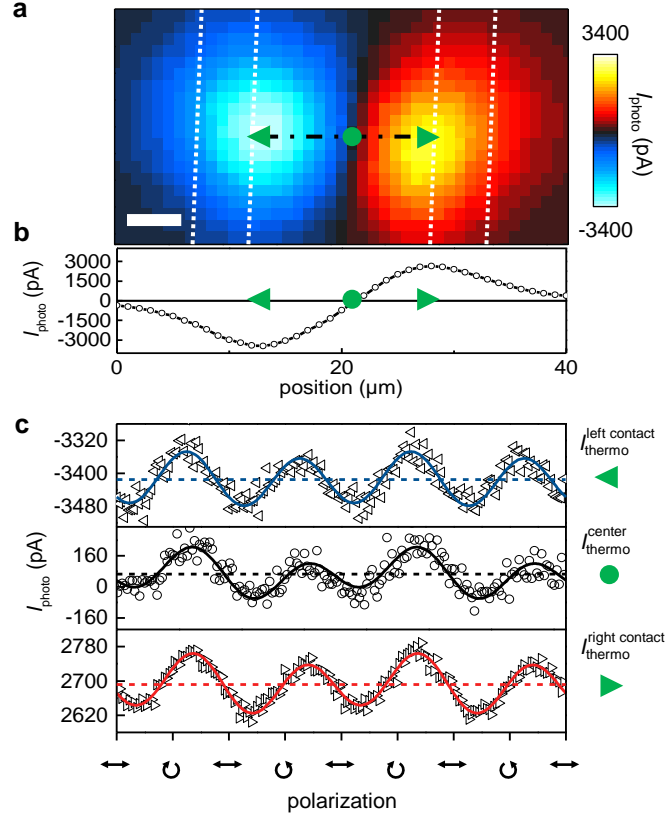

Fig. 2. Time-averaged photocurrents in $Bi_2Se_3$. (a) Photocurrent map of $I_{photo}$ in color scale. Dotted lines indicate the position of the striplines. Scale bar, 5 µm. (b) Line scan along the black dashed dotted line in Fig. 2a with a positive (rightward triangle) and a negative peak (leftward triangle) close to the striplines. Circle denotes position of zero signal of $I_{photo}$. (c) Polarization dependence of $I_{photo}$. Positions are denoted in Fig. 2a and b. The symbol ↔ denotes linearly polarized (along the $x$-axis), ↻ circularly right polarized, and ↺ circularly left polarized photons. The background contributions $I_{thermo}$ are denoted by dashed lines as described in Methods. Experimental parameters: 75 nm thin $Bi_2Se_3$-film, $E_{photon} = 1.53$ eV, $P_{laser} = 20$ mW, and $T_{bath} = 295$ K.

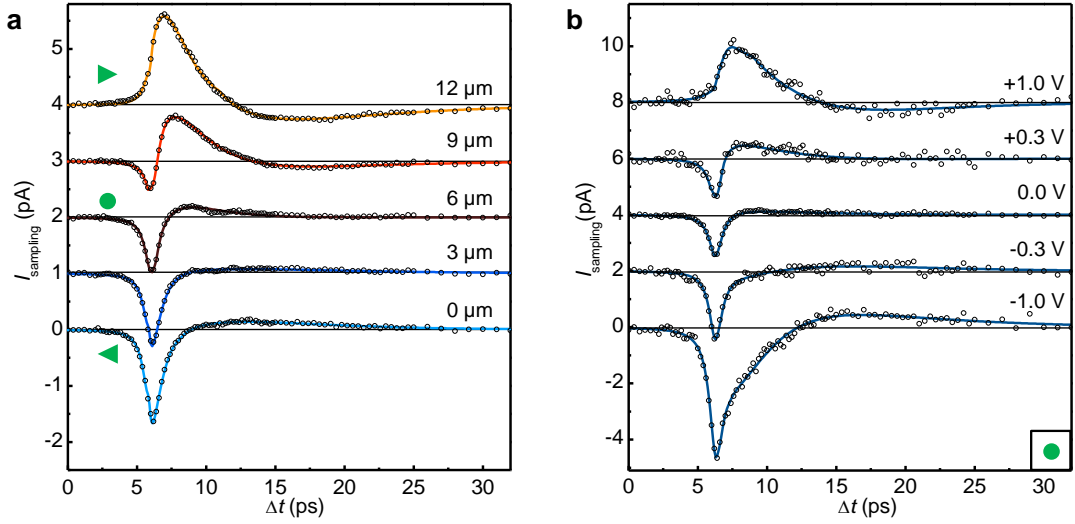

Fig. 3. Time-resolved expansion dynamics of hot electrons. (a) $I_{\text{sampling}}$ for excitation positions starting at the metal interface of a $Bi_2Se_3$-film in steps of 3 μm (from bottom to top) along the dashed dotted line in Fig. 2b. Triangles and circle are defined as in Fig. 2a. Solid lines are fits to the data according to Supplementary Note 2. Data are offset for clarity. Experimental parameters: 150 nm thin $Bi_2Se_3$-film, $E_{\text{photon}}$ = 1.59 eV, $P_{\text{laser}}$ = 1 mW, $V_{\text{sd}}$ = 0 V, $T_{\text{bath}}$ = 77 K, and a linear polarization. (b) $I_{\text{sampling}}$ for excitation position in the center of the $Bi_2Se_3$-film for -1 V ≤ $V_{\text{sd}}$ ≤ +1 V. The position is identical to the one as marked by a circle in Fig. 2a.

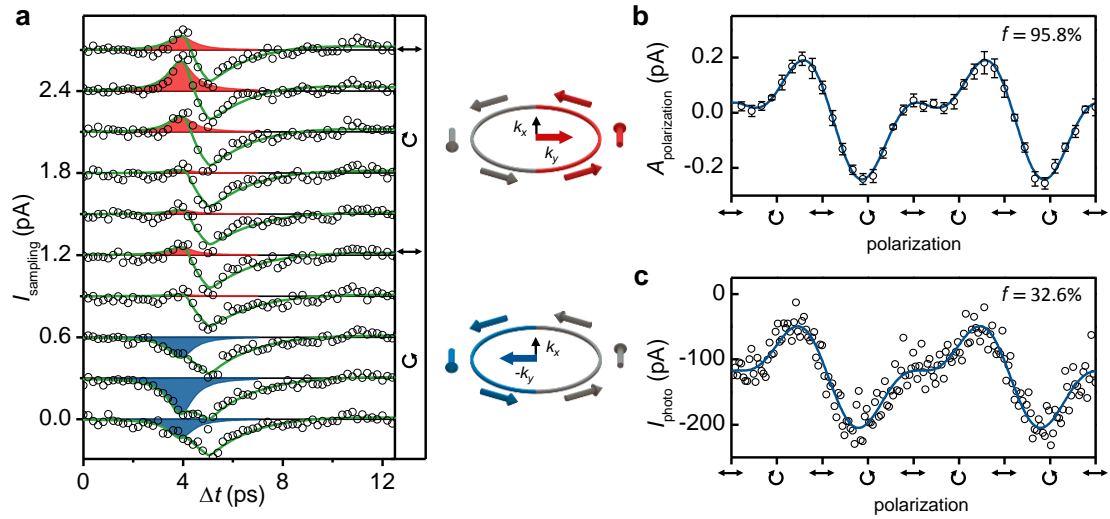

Fig. 4. Polarization control of ultrafast currents. (a) $I_{\text{sampling}}$ for excitation position close to the center of a $Bi_2Se_3$-film for varying polarization. The laser is focused at a position where the time-integrated photocurrent $I_{\text{photo}}$ is close to zero (circles in Fig. 2). Fits as described in Supplemental Note 2. Data are offset for clarity. Red and blue peaks are polarization controlled ultrafast currents in the direction of $k_y$ to the right contact (red) or $-k_y$ to the left contact (blue) with an in-plane spin polarization (middle panel). (b) Fitted amplitude of $I_{\text{sampling}}$ for $\Delta t = 4$ ps in Fig. 4a vs. photon-polarization. Error bars are fitting errors. (c) Simultaneously measured time-averaged $I_{\text{photo}}$ vs. photon-polarization. Experimental parameters: 90 nm thin $Bi_2Se_3$-film, $E_{\text{photon}} = 1.53$ eV, $P_{\text{laser}} = 20$ mW, and $T_{\text{bath}} = 295$ K.

# Supplementary Information

Ultrafast helicity control of surface currents in topological insulators with near-unity fidelity

Christoph Kastl, Christoph Karnetzky, Helmut Karl, Alexander W. Holleitner*

1. Polarization dependent photocurrents in $Bi_2Se_3$:

Microscopically, the circular photogalvanic effect can be understood by the following simplified symmetry argument. The pump laser excites the $Bi_2Se_3$-films under an oblique angle $\theta$ of incidence with a projection along the *x*-direction (Fig. 1b of the main manuscript). The helicity of the vector potential of the photons changes sign under a symmetry transformation with a mirror axis along this *x*-direction[1][2][3]. Only a current along the *y*-direction but not along the *x*-direction can follow this sign change (considering the symmetry $C_{3v}$ of the topological surface of $Bi_2Se_3$). In turn, the current of spin-polarized electrons to the contacts – as measured in our experiments – follows a sine curve as a function of the photon polarization[3]. We use this sinusoidal fingerprint to fit our data. In addition, the current is suppressed for a normal angle of incidence, as demonstrated below. The symmetry of the bulk states of $Bi_2Se_3$ is $D^5_{3d}$, which does not allow photogalvanic effects in the bulk[1][2][3].

Supplementary Figs. 1a, 1b, and 1c depict the polarization dependent photocurrents of a $Bi_2Se_3$ film under three angles $\theta = -17°$ (a), $0°$ (b), and $+17°$ (c) of (oblique) incidence at room temperature. Fitting the dependences with the sinusoidal fit function as given in the Methods section, we can extract a position dependence of the parameters *C*, $L_1$, $L_2$ and *D* (Supplementary Figs. 1d, 1e, and 1f). The parameters *C* and $L_1$ have opposite sign, as previously reported[4], and they vary qualitatively similar as a function of position. Furthermore, *C* and $L_1$ switch polarity for interchanging $\theta$ from -17° to +17° (d) and (f). Most importantly, the parameters *C* and $L_1$ are largely reduced for $\theta = 0°$ (e). Based on these observations, we interpret *C* to originate from the helicity-dependent circular photogalvanic effect within surface states with an in-plane helicity. $L_1$ stems from the helicity-independent linear photogalvanic effect, and it may comprise small contributions from the transverse photon drag effect[1]. As discussed below in the section "Energy levels in nanofabricated $Bi_2Se_3$ films", we consistently observe that *C* and $L_1$ occur at the same photon energies and therefore, optical transition energies within the $Bi_2Se_3$-films. Furthermore, for $\theta \neq 0°$, we attribute the components *D* and $L_2$ to originate from transitions in the bulk states of $Bi_2Se_3$. The above interpretations are in agreement with an earlier report[4].

We note that the range of $E_{photon} = 1.53 – 1.6$ eV used for the essential results in the main manuscript is chosen such that the contribution of the bulk and Rashba states can be considered to be negligible (see Supplementary Note 3 "Energy levels in nanofabricated $Bi_2Se_3$ films").

2. Time-of-flight analysis of the photogenerated hot carriers:

In the Supplementary Fig. 2a, the bottom trace of Fig. 3a of the main manuscript is reproduced (open circles). Assuming in a simplified model a time-scale separation of the different processes, we fit the data with the following function (green line):

$$I_{fit}(\Delta t) = I_{polarization}(\Delta t) + I_{right}(\Delta t) + I_{left}(\Delta t) \quad (S1).$$

The first term $I_{polarization}$ describes the polarization-dependent part of $I_{sampling}$ which can be fitted by a Lorentzian with the following form:

$$I_{polarization}(\Delta t) = A_{polarization} \cdot \sigma / [\pi(\Delta t^2 + \sigma^2)], \quad (S2)$$

with $A_{polarization}$ being the integrated area and $\sigma$ the half-width at half-maximum (HWHM) of the Lorentzian peak (gray in the Supplementary Fig. 2a). The Lorentzian considers phenomenologically the dispersion and attenuation of the electromagnetic transient running along the striplines before it is detected at the field probe. The two slower decaying components $I_{right}$ and $I_{left}$ have the form

$$I_i(\Delta t) = A_i/\delta_i^2 \cdot \Delta t \, e^{-\Delta t/\delta_i} \cdot \Theta(\Delta t), \quad (i = \text{right, left}) \quad (S3)$$

with $A_i$ being the corresponding integrated area and $\delta_i$ the characteristic decay time of $I_i$ ($i$ = right, left), and $\Theta(t)$ the Heaviside step function. $I_{right}$ describes the propagation of hot carriers to the right contact (red in the Supplementary Fig. 2a), and $I_{left}$ describes the propagation of hot carriers to the left contact (blue). We note that in Figs. 3b (zero bias) and in Fig. 4a of the main manuscript, experiments are presented in which the contributions of $I_{right}(\Delta t)$ and $I_{left}(\Delta t)$ are nearly negligible. It can be clearly seen that the residual peak fitted by $I_{polarization}(\Delta t)$ is a Lorentzian and not a Gaussian.

We observe that $I_{polarization}$ is the only term in the time-resolved photocurrent $I_{sampling}$ which depends on the photon polarization (Fig. 4a of the main manuscript). In addition, this term vanishes on a time scale which is consistent with the spin-depolarization time in $Bi_2Se_3$ convoluted with the time-resolution of the presented experiment (~1ps)[4]. We point out that the striplines act as nearfield antennas in the THz-regime. Hereby, the striplines pick up the electromagnetic transients produced by the photocurrents quasi instantaneously via the speed of light[5]. In turn, the first peak term $I_{polarization}$ defines the time $t_0$ when the pump laser excites the $Bi_2Se_3$-films. $I_{right}$ and $I_{left}$ describe slower processes of hot charge carries propagating within the $Bi_2Se_3$-films. Consistently, both $I_{right}$ and $I_{left}$ depend on the excitation position with respect to the $Bi_2Se_3$-films and on $V_{sd}$ (Fig. 3 of the main manuscript). We note that the simplifying fit of equation (S1) partially does not comprise the situation when polarization-dependent and polarization-independent optoelectronic processes somewhat overlay. For instance, this can happen at the contacts (such as top trace of Fig. 3a) and at the presence of electric fields (top trace of Fig. 3b). That is the reason why all helicity-dependent currents are only discussed for the center of the $Bi_2Se_3$-films and for a bias close to zero (as in Figure 4).

With respect to $t_0$, the maxima of $I_{right}$ ($I_{left}$) occur at a delayed time $t_0^{right}$ ($t_0^{left}$) (Supplementary Fig. 2a) depending on the distance to the contacts. The Supplementary Fig. 2b represents a corresponding time-of-flight diagram of the 150 nm thin $Bi_2Se_3$-film at 77 K which allows us to estimate an upper limit of the propagation velocity of the fastest photogenerated electrons to be $v_e = (6.1 \pm 0.6) \times 10^5$ ms$^{-1}$. For the 90 nm thin $Bi_2Se_3$-film at 295 K (Supplementary Fig. 3), we extract $v_e = (5.7 \pm 0.7) \times 10^5$ ms$^{-1}$. Both values are consistent with the group velocity of $(5.5 \pm 0.2) \times 10^5$ ms$^{-1}$ as calculated in the section "Discussion" of the main manuscript. We point out that the maxima of $I_{right}$ and $I_{left}$ comprise the fastest propagating photogenerated electrons. The decaying tails of the $I_{right}$ and $I_{left}$ comprise slower photocurrent dynamics with a time scale of several picoseconds. The underlying processes are a combination of diffusion and drift of hot electrons due to the thermopower and electrostatic potentials. The latter scenario is proven by the fact that $t_0^{right}$ and $t_0^{left}$ depend on $V_{sd}$ at the middle of the $Bi_2Se_3$-films (Suppl. Fig. 2c).

The bias dependence further allows us to conclude that the time-resolved photocurrents are dominated by photogenerated electrons and not holes. This will be discussed in the following. In Supplementary Fig. 4 we depict schematically the photocurrents of electrons and holes for different excitation positions and under different external bias conditions. The arrows show the currents (Supp. Fig. 4a), whereby the direction of the arrows indicates the current direction and the length indicates the time of flight towards the contact. In the Supplementary Fig. 4b, we depict the currents for excitation at the left contact. The currents due to charge carriers propagating to the right are delayed compared to currents to the left, as indicated. The situation is reversed in Fig. 4d. For excitation exactly at the middle (Supp. Fig. 4c) all currents are equally delayed and cancel out. In the Supplementary Figs. 4e and 4f, the electric field shifts the transport currents. Note the asymmetry between electrons and holes. Experimentally we find in Fig. 3 of the main manuscript, that the experimental condition in the Supplementary Figs. 4b and 4e yield qualitatively equal time-resolved photoresponses. Equally, supplementary Fig. 4d give the equivalent results as Fig. 4f. Both consistencies are only possible, when the transport is dominated by electrons and not by holes, as indicated by the crossed-out contribution of the hole currents in the Supplementary Figs. 4e and 4f.

3. Energy levels in nanofabricated Bi$_2$Se$_3$ films:

Supplementary Fig. 5a sketches the band diagram for *n*-type Bi$_2$Se$_3$ without band bending effects at the surface, as was experimentally derived by two independent experiments using a two-photon angle resolved photoemission spectroscopy (2PPES)[10][11]. A very similar band diagram was theoretically computed by density functional theory[10][11][12]. Generally, the relative energy differences of the bulk bands and the surface states depend on the surface treatment of the Bi$_2$Se$_3$ films; i.e. the exposure to air as is essential for nanofabricated Bi$_2$Se$_3$ films with metal contacts. For all treatments, Bi$_2$Se$_3$ ends up to be *n*-type with a Fermi energy of several 100s of meV, and the transition energies can additionally vary by ~0.1 eV.[13][14][15] Importantly, the coexistence of the topological surface states to such a surface inversion layer was experimentally verified.[15][16]

We fabricated a Bi$_2$Se$_3$-film with a height of 65 nm from the same batch of materials with four metal contacts to perform Hall-measurements. We obtain an electron density of ~4·10$^{19}$ cm$^{-3}$ and an electron mobility of 470 cm²/Vs at room temperature. Accordingly, we obtain a Fermi-energy of ~0.3 eV consistent with earlier reports. The depth of the inversion layer can be estimated via the Thomas-Fermi screening length to be ~6 nm, consistent with earlier reports[13][15][16].

To perform 2PPES on nanofabricated Bi$_2$Se$_3$ films is a nearly impossible task. However, the results by refs. [10][11][12] verify for *n*-type Bi$_2$Se$_3$ that a photon with energy of ~1.5 eV excites charge carriers from the first conduction band (CB$_1$) to an unoccupied, topological non-trivial Dirac cone of the surface state SS$_2$ (orange arrow and CB$_1$ and SS$_2$ as in the Supplementary Fig. 5a). This energy is slightly less than the photon energy used in our optoelectronic experiments. Therefore, we exploit the photocurrent signal itself to conclude which surface states are involved in the polarization dependent currents.

Supplementary Figs. 5b – 5d depict the (polarization dependent) photocurrent $I_{photo}$ vs. $E_{photon}$. In particular, for each $E_{photon}$, a photocurrent map is measured, and the corresponding maximum amplitude of $|I_{photo}|$ of the maps is extracted vs. $E_{photon}$ (Supplementary Figure 5b). Then, at the center position of the maps, the polarization dependent photocurrent is measured and analyzed. Supplementary Fig. 5b (c,d) depicts the corresponding parameter *C* ($L_1$, $L_2$) for a laser energy in the range of 1.33 eV ≤ $E_{photon}$ ≤ 1.66 eV. We find that the parameter *C* and $L_1$ show a very similar energy dependence throughout the examined range of $E_{photon}$. This corroborates the interpretation that they stem from related optoelectronic processes – the circular and the linear photogalvanic effects from surface states (see section "Polarization dependent photocurrents in Bi$_2$Se$_3$"). The correlation between *C* and $L_1$ has been experimentally verified by earlier work[4].

The dispersion of the states SS2 and CB1 are such that the transition CB$_1$ → SS$_2$ has a maximum energy at a non-zero *k*-vector ≤ 0.1 Å$^{-1}$ (orange arrow in the Supplementary Fig. 5a)[10] and for all other *k*-vectors, the transition energy is smaller. For an *n*-type inversion layer with a Rashba spin-split CB1, the Rashba spin-split bands are ~0.13 eV above a non-split CB1 [13][15]. Therefore, the possible transition energy $CB_1 \rightarrow SS_2$ at the surface is reduced by about ~0.1 eV. In turn, we can assume for a photon energy $E_{photon} \geq$ ~1.5 eV, the transition $CB_1 \rightarrow SS_2$ is not relevant at the surface. In the bulk, the transition cannot occur either because there, the surface states do not exist.

We note that $L_2$ exhibits a maximum at 1.45 eV $\leq E_{photon} \leq$ 1.53 eV (supplementary Fig. 5e), which we attribute to a dominant transition $VB_1 \rightarrow VB_2$ in this energy range (supplementary Figure 5a). Hereby, we attribute $L_2$ to originate from bulk transitions in $Bi_2Se_3$. This is again in agreement with conclusions based on earlier optoelectronic experiments[4]. Consistently, the maximum photocurrent $|I_{photo}|$ as depicted in the supplementary Fig. 5a varies in a similar manner as $L_2$ for 1.45 eV $\leq E_{photon} \leq$ 1.53 eV. As recently demonstrated[13], the valence band $VB_1$ at the surface is not spin-split. Therefore, the transition $VB_1 \rightarrow SS_2$ does not contribute to the circular photogalvanic effect. Last but not least, the bulk transition $CB_1 \rightarrow CB_2$ can only be expected for a photon energy $E_{photon} \geq$ 1.7 eV [10][12].

To summarize, we chose $E_{photon}$ = 1.53 eV for the presented time-resolved, polarization-dependent experiments (Fig. 4a in main manusscript), because $L_2$ exhibits a minimum, and C and $L_1$ are in a maximum range. Based on the above arguments, we optically excite the transitions $SS_1 \rightarrow SS_2$. This interpretation is consistent with the fact that we find a propagation velocity of $(5.5 \pm 0.2) \times 10^5$ ms$^{-1}$ at room temperature (see Supplementary Note 2 "Time-of-flight analysis of the photogenerated hot carriers").

We point out that the chosen range of $E_{photon}$ is in the range 1.5 eV $< E_{photon} \leq$ 1.7 eV, in which we can argue that the transition $SS_1 \rightarrow SS_2$ is responsible for the circular photogalvanic effect. However, the revealed ultrafast optoelectronic dynamics of surface states in topological insulators applies throughout the whole vis-NIR range, since transition $SS_1 \rightarrow SS_2$ is always excited in this range.

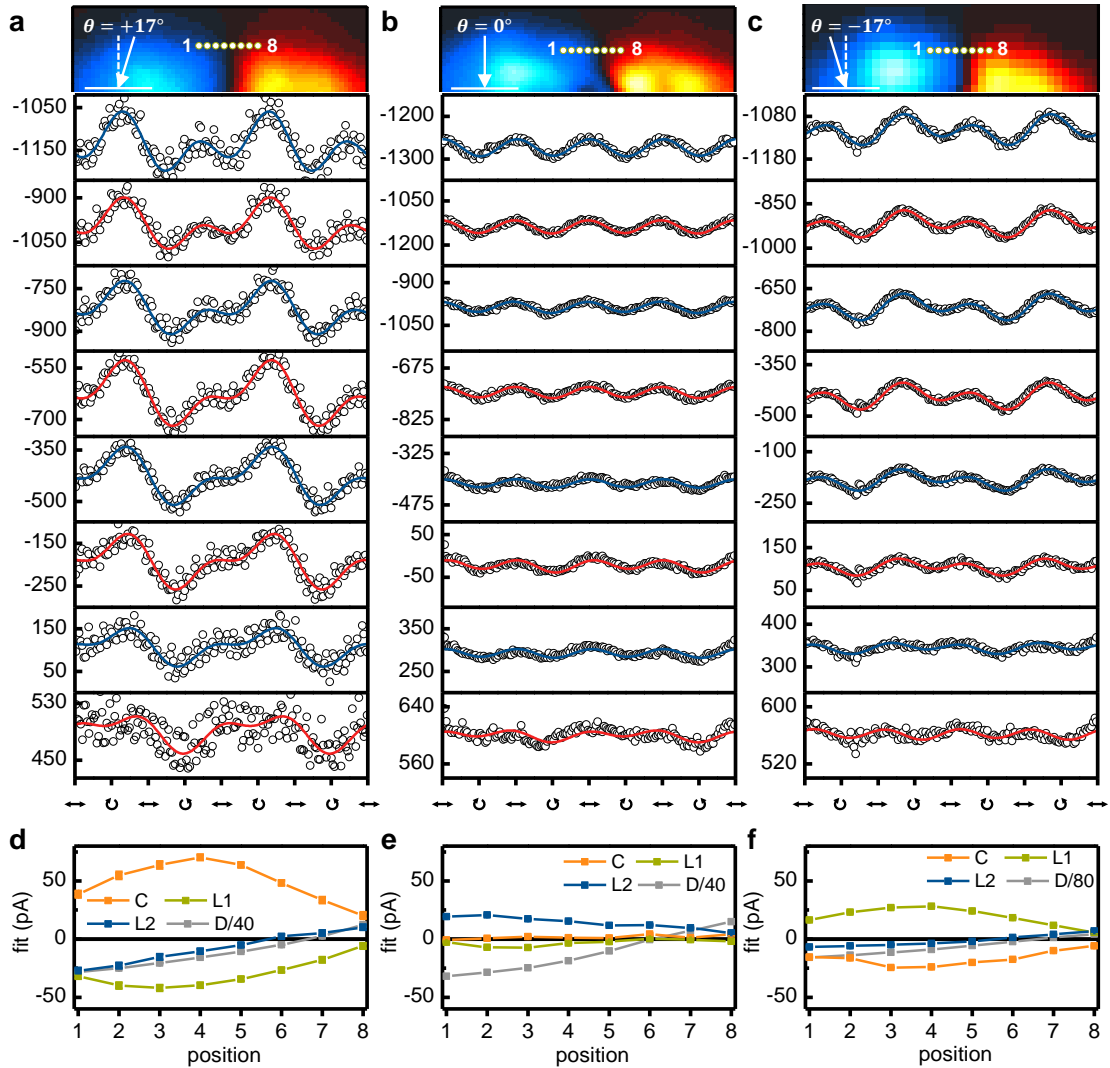

**Supplementary Fig. 1 Polarization dependence of the photocurrent $I_{photo}$ for a varying oblique angle $\theta$. a**, **b**, and **c**, $\theta$ = -17°, 0° and +17° as defined in the top panels. The top panels also display photocurrent maps (as in Fig. 2a of the main manuscript). The dots in the maps define the eight positions 1, … to 8, at which the polarization dependence of $I_{photo}$ are measured. The eight polarization dependences of $I_{photo}$ are depicted below the photocurrent maps, respectively. The experimental parameters: 90 nm thin $Bi_2Se_3$-film, $E_{photon}$ = 1.53 eV, $P_{laser}$ = 10 mW, and $T_{bath}$ = 295 K. Lines are sinusoidal fits as discussed in the Methods section. Figures **d**, **e**, and **f**, depict the corresponding fitting parameters. Most importantly, $C$, $L_1$ change their current polarity for $\theta$ = -17° and +17°, and they are close zero for $\theta$ = 0°. For clarity, $D$ is scaled down as specified.

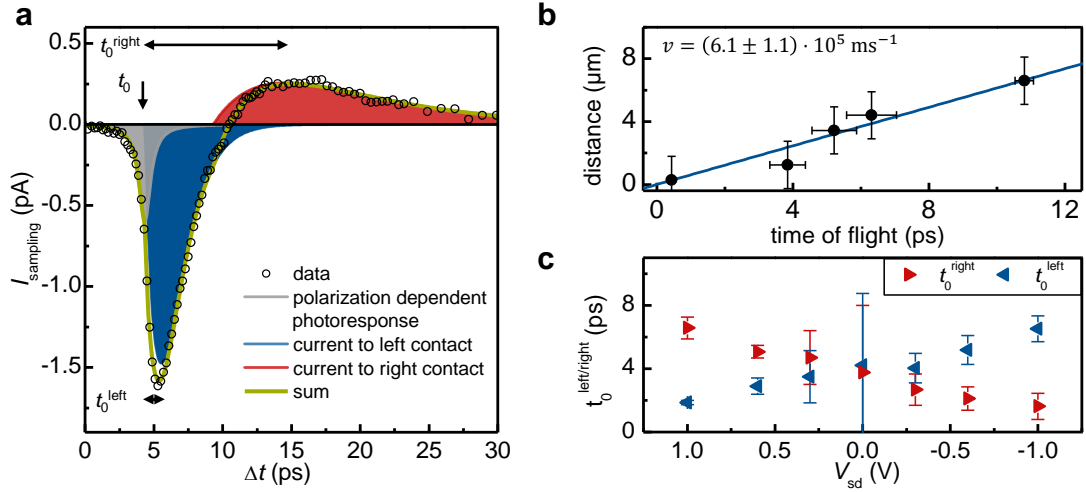

**Supplementary Fig. 2 Time-of-flight analysis of photogenerated hot carriers the Bi$_2$Se$_3$-film with thickness of 150 nm at 77 K. a**, $I_{\text{sampling}}$ as a function of the time delay $\Delta t$. This graph introduces the fitting functions and the time-delays at which the current to the right contact ($t_0^{\text{right}}$) and the one to the left contact ($t_0^{\text{left}}$) currents are maximum. **b**, Distance of propagation vs. time of flight of the photogenerated electrons. **c**, Voltage dependence of $t_0^{\text{right}}$ and $t_0^{\text{left}}$ for $|V_{\text{sd}}| \leq 1$ V at the middle of the Bi$_2$Se$_3$ film. The experimental parameters: 150 nm thin Bi$_2$Se$_3$-film, $E_{\text{photon}} = 1.59$ eV, $P_{\text{laser}} = 1$ mW, and $T_{\text{bath}} = 77$ K.

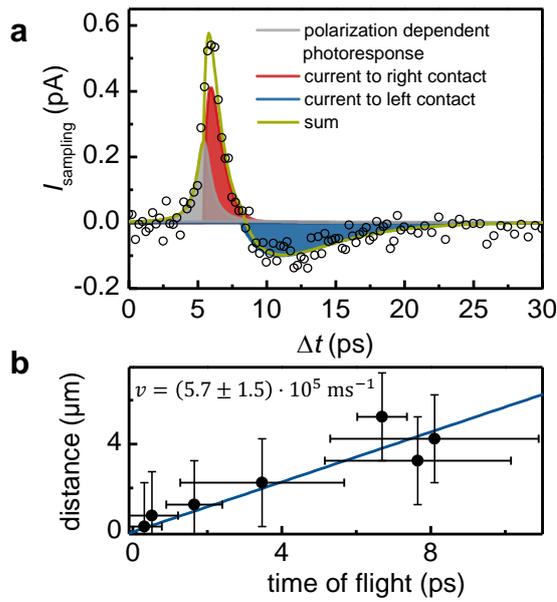

**Supplementary Fig. 3 Time-of-flight analysis of photogenerated hot carriers in the Bi$_2$Se$_3$-film with thickness of 90 nm at room temperature. a**, $I_{sampling}$ as a function of the time delay $\Delta t$ with fitting functions. **b**, Distance of propagation vs. time of flight of the photogenerated electrons. The experimental parameters: 90 nm thin Bi$_2$Se$_3$-film, $E_{photon}$ = 1.53 eV, $P_{laser}$ = 20 mW, and $T_{bath}$ = 295 K.

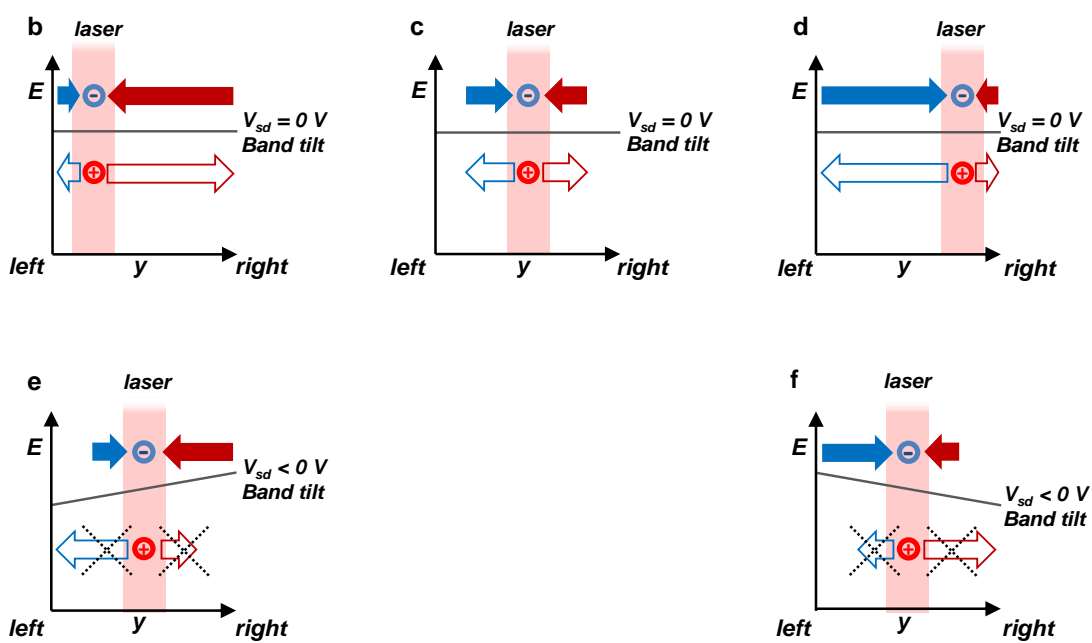

**Supplementary Fig. 4 Schematic of the expansion direction of photogenerated electrons and holes.** See Supplementary Note 2 for details.

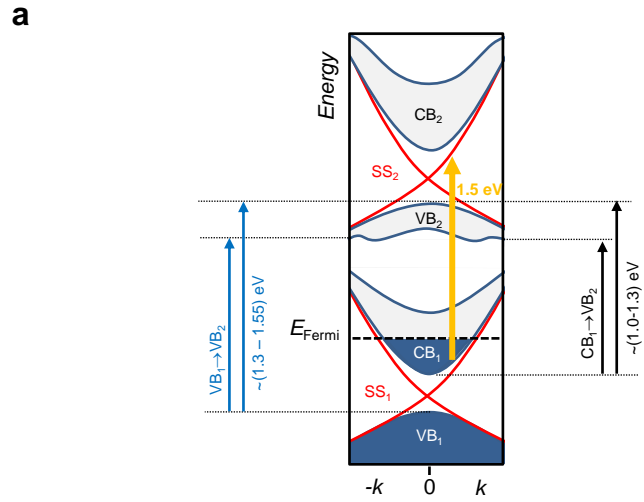
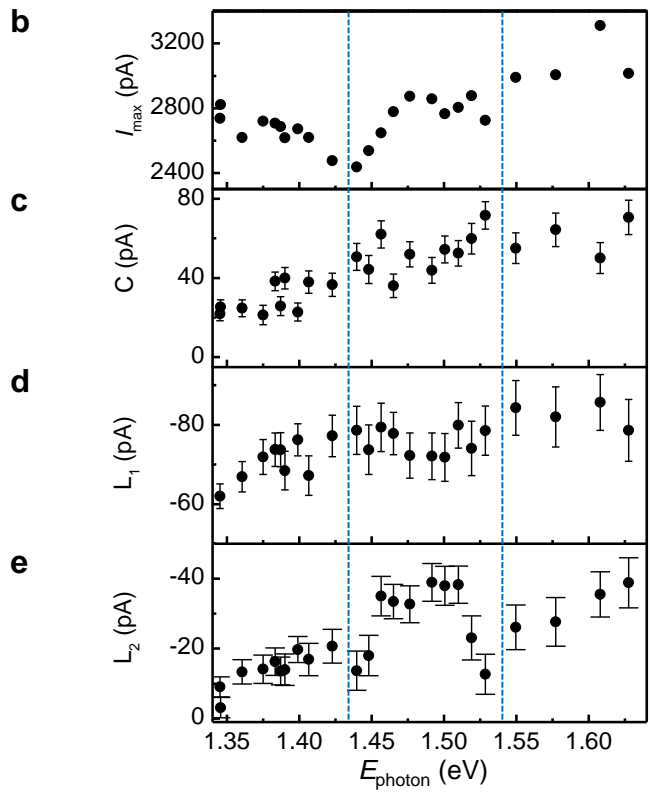

**Supplementary Fig. 5 Polarization dependent photocurrents vs. photon energy.**
**a**, Sketch of the band diagram of $Bi_2Se_3$ based on ref. [30]. **b**, Maximum of $|I_{photo}|$ vs. $E_{photon}$. **c**, Parameter $C$ vs. $E_{photon}$. **d**, Parameter $L_1$ vs. $E_{photon}$. **e**, Parameter $L_2$ vs. $E_{photon}$. The experimental parameters: 75 nm thin $Bi_2Se_3$-film, $P_{laser}$ = 20 mW, and $T_{bath}$ = 295 K.